# NESNE YÖNELİMLİ YAZILIM KALİTESİ AÇISINDAN C VE K METRİK KÜMESİNİ DEĞERLENDİREN BİR UZMAN MODÜL TASARIMI VE UYGULAMASI

# *(AN EXPERT MODULE DESIGN AND IMPLEMENTATION THAT EVALUATION THE C&K METRIC SUITE IN TERMS OF OBJECT ORIENTED SOFTWARE QUALITY)*

**M.Hanefi CALP[*], Nursal ARICI[**]**

## ÖZET/*ABSTRACT*

Nesne yönelimli yazılımlar kapsülleme, kalıtım, uyumluluk, bağımlılık ve çok biçimlilik gibi bazı özelliklere sahiptir. Bu özellikler, nesne yönelimli yazılımların test, bakım ve onarım faaliyetlerini önemli ölçüde etkilemektedir. Söz konusu faaliyetlerin başarılı bir şekilde sonuçlanması, değerlendirilecek metriklerin doğru teşhis edilmesine ve doğru yorumlanmasına bağlıdır. Bu metriklerin güvenilir ve hızlı bir şekilde değerlendirilmesi de ancak yazılım test araçlarının geliştirilmesi ve kullanılmasıyla mümkündür. Günümüzde nesne yönelimli metrikler için birçok yazılım test aracı mevcuttur. Ancak, bu araçların yetenekleri konusunda birtakım sınırlılıklar bulunmaktadır. Sözkonusu araçların birçoğu sadece metrik bulgularını üretmekte, bu bulguların ne anlama geldiği ve yazılım kalitesini nasıl etkilediği noktasında herhangi bir işlem yapmamaktadır. Dolayısıyla, bu araçların çeşitlendirilmesi ve yeteneklerinin arttırılması gerekmektedir. Yazılım test araçlarının sözü edilen sınırlılıklarından yola çıkarak bu çalışma ile, nesne yönelimli test sürecine uzman sistem yaklaşımı ile otomasyon sağlamak amaçlanmıştır. Bu amaç doğrultusunda, nesne yönelimli programlama metodolojisine göre geliştirilen bir projenin (jar uzantılı) iç kalite özellikleri hakkında bilgi veren metrik bulguları yorumlayan bir uzman modül gerçekleştirilmiştir. Bu modülde, Chidamber ve Kemerer (CK)'in geliştirdiği metrik kümesi kullanılmıştır. Metrics adlı yazılım test aracı vasıtasıyla elde edilen CK metrik bulguları otomatik, el ile ve sürümler arası karşılaştırma yöntemleri ile değerlendirme ve iç kalite özellikleriyle ilişkisini yorumlama özelliğine sahiptir. Sözkonusu modülün yazılım test faaliyetlerinde geçerli olan beyaz kutu testlerine bir destek unsuru olması beklenmektedir.

*Object-oriented software has some features such as encapsulation, inheritance, cohesion, coupling and polymorphism. These features significantly affect the testing, maintenance and repair activities of object-oriented softwares. The successfully conclusion of these activities, depends on correct detect and correct interpretation of metrics will be evaluated. Also safely and rapidly evaluation of this metrics, is possible but by developing and using the software testing tools. Nowadays, many software testing tools that there are for object-oriented metrics. However, there are some limitations on the capabilities of this tools. Many of the this tools produces only metric findings, does not take any action in the point what this findings mean and how it affects the quality of the software. Therefore, diversifying and increasing capabilities of these tools is required. Based on the limitations of software testing tools mentioned in this study, is intended to provide the automation by expert system approach to object-oriented testing process. For this purpose, an expert module that review and interpret the metric findings, that providing information about the internal quality attributes of a project (.jar extension) developed according to object-oriented programming methodology have been developed. Metric suite that developed by Chidamber and Kemerer (C&K) have been used in this modul. This module, metric findings obtained by software testing tool named Metrics, has the evaluation feature by methods automatic, manuel and cross-version comparison and the interpreting the relationship between internal quality specifications. It has been expected that the betreffend module to be a support element to white-box tests been valid in software testing activities.*

* Gazi Üniversitesi, Bilişim Enstitüsü, ANKARA
** Gazi Üniversitesi, Teknoloji Fakültesi, Bilgisayar Mühendisliği, ANKARA

## 1. GİRİŞ

Nesne yönelimli yazılım tasarımı ve kodlanması, yazılım geliştirmenin popüler bir yolu olmuştur. Dolayısıyla, bu yazılımların test, bakım ve onarım faaliyetleri de yapısal programlamaya göre farklılıklar göstermektedir. Bu faaliyetlerin başarılı bir şekilde sonuçlanması, değerlendirilecek metriklerin doğru teşhis edilmesine ve doğru yorumlanmasına bağlıdır (Pressman, 2005).

Metrikler, yazılım kalitesinin ölçülmesi ve geliştirilmesi faaliyetlerinde önemli bir yer tutmaktadır. Bu faaliyetler; aşırı bağımlı, uyumsuz, karmaşık ve hatalı modüllerin belirlenmesini sağlar. Sözkonusu belirlemeler, hangi modüllerin öncelikli olarak test edileceği veya hangi testlerin öncelikli olarak uygulanacağı, bakım-onarım için ne kadar bütçe ve zaman harcanacağı gibi bilgilere ulaşmada önemli ipuçları verir (Erdemir vd., 2008).

Literatürde nesne yönelimli tasarım metriklerin önemi, özellikleri, görevleri, tasarım kalitesine etkisi ve bu metrikler arasındaki ilişkiler ile ilgili birçok çalışma bulunmaktadır (Lanza ve Marinescu, 2006; Lanza ve Ducasse, 2003; Chidamber ve Kemerer, 1994; Lanza ve Ducasse, 2001; Sellers, 1996; Laing ve Coleman, 2001; Bellin vd., 1999; Sarker, 2005 Robert, 1995; Baroni vd., 2002). Bununla birlikte, tasarım metrikleri kullanılarak yazılım kalite değerlendirmesi yapan çalışmalar da mevcuttur (Marinescu, 2004; Çatal ve Diri, 2008; Brito vd., 1996; Harrison ve Counsell, 1998; Basili vd., 1996). Ayrıca, yazılım bakımı ile ilgili yapılan çalışmalar ve uzman sistem kullanarak nesne yönelimli metriklerin davranışlarını anlamayı amaçlayan çalışmalar da yapılmıştır (Sheldon vd., 2002; Chapin vd., 2001; Tao, 1995).

Yapılan bu çalışmalarda, metrikleri ve bu metriklerin birbirleriyle olan ilişkilerini anlamanın, yazılım tasarım kalitesine katkı sağlayacağı açıkça vurgulanmıştır. Aynı zamanda metriklerin; nesne yönelimli yazılımların yapısındaki bağımlılık, kalıtım ve karmaşıklık gibi tasarım kaliteleri hakkında önemli ipuçları verdiği ortaya konulmuştur. Çalışmalarda özellikle vurgulanan diğer bir nokta da, yazılım tasarım kalitesini ölçmek için metriklerin tek başına değil, bir arada kullanılması gerekliliğidir. Aksi takdirde gerçekçi veya doğru sonuçlara ulaşmak mümkün olmayacaktır (Lanza, 2006; Lanza ve Ducasse, 2003; Chidamber ve Kemerer, 1994; Lanza ve Ducasse, 2001; Harrison ve Counsell, 1998).

Bu çalışmada, ikinci bölümde nesne yönelimli yazılım metrik kümesi ve önemi, üçüncü bölümde geliştirilen Chidamber ve Kemerer metrik değerlendirme modülünün (CKMDM) tasarımı ve uygulaması, son olarak dördüncü bölümde ise tüm bu çalışmalardan çıkarılan sonuç ve öneriler yer almaktadır.

## 2. NESNE YÖNELİMLİ YAZILIM METRİK KÜMESİ VE ÖNEMİ

Nesne yönelimli metrikler, yazılım test sürecinin etkililiğinin önemli bir göstergesidir. Sözkonusu metrikler proje yöneticilerinin, metriklerin durumlarını anlamak ve yazılım teslimi ile ilgili riskleri azaltmak için kullanabildiği anahtar olaylardır. Metrikler, hâlihazırdaki performansı ölçmede ve yazılım projelerini daha iyi kontrol etmede önemli rol oynarlar. Aynı zamanda ürünün fonksiyonları hakkında daha fazla bilgi edinmeyi ve projelerin gelecekteki kalitesini daha iyi tahmin etmeyi sağlarlar (Pusala, 2005).

Literatürde CK, Brito (Metrics for Object Oriented Design-MOOD) ve Bansiya ve Davis QMOOD (Quality Model for Object-Oriented Design) gibi nesne yönelimli yazılım metrik kümeleri tanımlanmış olup, bu çalışma literatürde yaygın olarak kabul gören CK'nın metrik kümesinin değerlendirilmesi ile sınırlı tutulmuştur (Erdemir vd., 2008; Calp, 2011). CK metrik kümesinin seçilmesinin sebebi, hem daha çok nesne yönelimli yazılım özelliklerini (kapsülleme, kalıtım, uyum vb.) kapsaması hem de bir sistemin bütün olarak değerlendirilmesinden ziyade sistemdeki sınıfları geniş bir şekilde değerlendirme imkânı tanımasıdır. Aynı zamanda, bu sınıfların karmaşıklık, verimlilik, anlaşılırlık, test edilebilirlik ve dayanıklılık gibi iç kalite özelliklerini ölçmede önemli ölçüde yardımcı olması sözkonusu metrik kümesinin seçilmesinde etkili olmuştur. C&K, nesne yönelimli tasarım için altı adet metrik tanımlar. Bu metrikler ve özellikleri kısaca şöyledir:

*Sınıfın ağırlıklı metot sayısı (SAMS) (Weighted Methods per Class):* Bir sınıftaki metotların karmaşıklık derecesi veya sayısıdır.

*Kalıtım ağacının derinliği (KAD) (Depth of Inheritance Tree):* Sınıfın kalıtım ağacının köküne uzaklığıdır.

*Alt sınıf sayısı (ASS) (Number of Childrens):* Bir sınıftan direk türetilmiş alt sınıfların sayısıdır.

*Nesne sınıfları arasındaki bağımlılık (NSB) (Coupling Between Object Classes):* Sınıflar arasındaki bağımlılık, bir sınıf içindeki özellik (attribute) ya da metotların (method) diğer sınıfta kullanılması ve sınıflar arasında kalıtımın olmaması durumudur.

*Sınıfın tetiklediği metot sayısı (STMS) (Response For a Class):* Bir sınıftan bir nesnenin metotları çağrılması durumunda, bu nesnenin tetikleyebileceği tüm metotların sayısıdır. Yani, bir sınıfta yazılan ve çağırılan toplam metot sayısıdır.

*Metotlardaki uyum eksikliği (MUE) (Lack of Cohesion in Methods):* MUE, n adet kümenin kesişiminden oluşan kümelerdeki uyumsuzlukların sayısıdır ve metotlardaki benzerlik derecesidir (Erdemir vd., 2008; Chidamber ve Kemerer, 1994; Calp, 2011a; Chidamber ve Kemerer, 1991).

## 3. GELİŞTİRİLEN UZMAN MODÜLÜN TASARIMI VE UYGULAMASI

Geliştirilen veya tanımlanan metriklerin yorumlanması noktasında literatürde bazı yöntemler mevcuttur. Bunlardan biri, *"özellik tabanlı metrik ölçümü"*, diğeri *"belirlenen metriklerin ürettiği değerler ile aynı amaca hizmet eden diğer metriklerin ürettiği değerler arasındaki ilişkinin ölçülmesi"*, bir diğeri de "*metrik grubunun ölçülecek olan yazılımın sürümleri üzerinde uygulanması"*dır. İlk yöntemde, sözkonusu metriklerin özellikleri dikkate alınarak değerlendirme faaliyetleri gerçekleşmektedir. Örneğin, geliştirilen yazılımdaki kod sayısı veya alt sınıf sayısı o yazılımın karmaşıklığı hakkında bilgi sağlamaktadır. İkinci yöntemde, belirlenen metriklerin ürettiği değerler ile aynı amaca hizmet eden diğer metriklerin ürettiği değerler arasındaki ilişki ölçülmektedir. Son yöntemde ise, önceki yazılım sürüm/lerin metrik değerleri ve kusur bilgileri kullanılarak bir model ortaya çıkarılmaktadır. Ardından yeni sürümdeki metrik değerleri ve kusur bilgileri ile karşılaştırılarak kalite düzeyi yorumlanmaktadır. Bu metrik değerlerini ve kusur bilgilerini yorumlamada genetik algoritmalar, yapay sinir ağları, uzman sistemler veya karar ağaçları gibi farklı yöntemleri kullanan test araçları bulunmaktadır (Erdemir vd., 2008; Çatal ve Diri, 2008a; Calp, 2011; Briand vd., 1996; Çatal ve Diri, 2007; Çatal ve Diri, 2008b). Günümüzde de Findbugs, Metrics, PMD ve Coverity gibi ücretli veya ücretsiz nesne yönelimli yazılım test araçları geliştirilmiştir (Findbugs, 2012; Metrics, 2012; PMD, 2012; Coverity, 2012). Ancak, bu araçlar genellikle metrik değerlerine ilişkin bulgular/sayısal değerler vermekle sınırlı kalmaktadır. Sözkonusu sınırlılıklardan yola çıkarak bu çalışmada uzman sistemler yaklaşımıyla C&K'in metrik kümesini değerlendiren bir uzman modül gerçekleştirilmiştir. Geliştirilen uzman modül, elde edilen sayısal metrik bulguları hem otomatik hem el ile hem de sürümler arası değerlendirme yöntemi ile yorumlamaktadır. Yorumlama işlemi, *özellik tabanlı metrik ölçümü* ve *metrik bulgularının sürümler arası karşılaştırılması* yöntemleri kullanılarak yapılmaktadır. Aşağıdaki kesimlerde, geliştirilen uzman modülün yapısı, uygulama adımları ve işleyişi ayrıntılı olarak açıklanmıştır.

### 3.1. Geliştirilen Uzman Modülün Yapısı

Uzman sistemler (US), belirli bir konuda uzman olan bir veya birçok insanın yapabildiği muhakeme ve karar verme işlemlerini modelleyen bir yazılım sistemidir (Nabiyev, 2005). Yazılım mühendisliği literatüründe uzman sistemler kullanılarak yazılım analizi, test durumlarını seçme ve test planı oluşturma, yazılım risk analizi yapma, yazılım mühendisliği yönetimi ve nesne yönelimli davranışları durum değişikliklerine göre değerlendirme gibi birtakım çalışmalar yapılmıştır (Muratore ve Demasie, 1991; Xu vd., 2005; Xu vd., 2003; Ramsey ve Basili, 1986; (Tao, 1995). Ancak, nesne yönelimli tasarım özelliklerinin ele alınması ve değerlendirilmesi ile ilgili birçok eksik bulunmaktadır. Sözkonusu eksiklikler bu çalışmanın hazırlanmasına sebep olmuştur. Şekil 1'de geliştirilen uzman modülün genel yapısı açıkça gösterilmiştir.

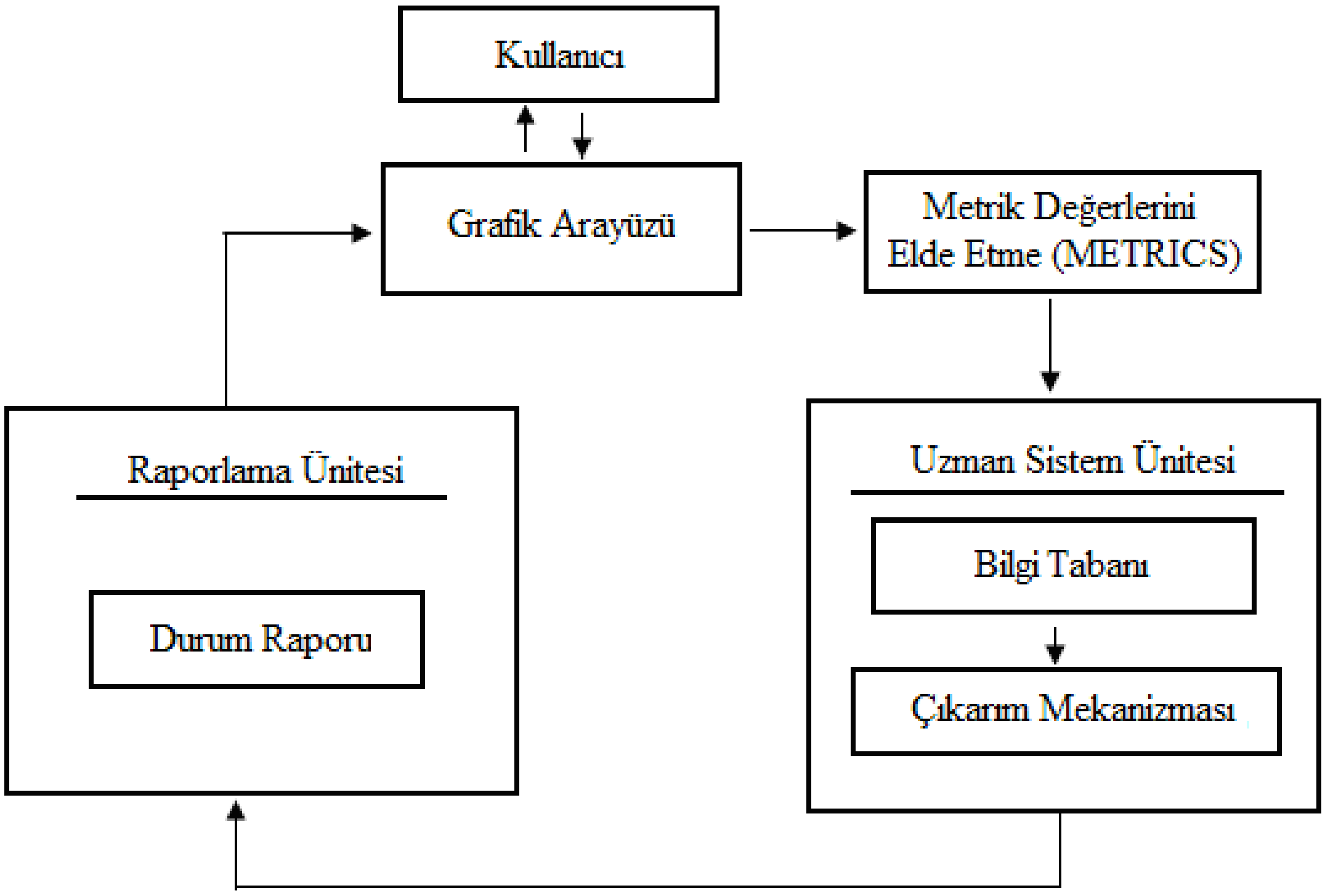

Şekil 1. Geliştirilen uzman modülün yapısı

<u>Kullanıcı</u>

Uzman modülün kullanıcı kısmında; yazılım geliştiricileri, test mühendisleri, test yöneticileri, proje yöneticileri gibi yazılım proje ekibinde yer alan kişiler olabilmektedir.

<u>Grafik Arayüzü</u>

Bu bölüm, kullanıcılar ile sistem arasındaki iletişimin sağlandığı bölümdür. Kullanıcılar tarafından grafik arayüzü kullanılarak sisteme giriş yapılabilmekte, istenilen durumda daha önce oluşturulmuş metrik bulguları kullanılabilmekte veya yeni metrik bulguları oluşturulabilmektedir. Elde edilen bulgular değerlendirilerek sistem tarafından üretilen raporlara erişilebilmekte ve bu raporlar bilgisayar bellek birimlerine kaydedilebilmektedir.

<u>Metrik Değerlerini Elde Etme</u>

Kullanıcı, uzman modüle entegre edilen "Metrics" (Yıldız Teknik Üniversitesi Bilgisayar Mühendisliği, 2012) adlı programı kullanarak değerlendirme yapmak için gerekli olan C&K'in metrik bulgularını hem sınıf hem de proje bazında elde edebilmektedir (Şekil 2).

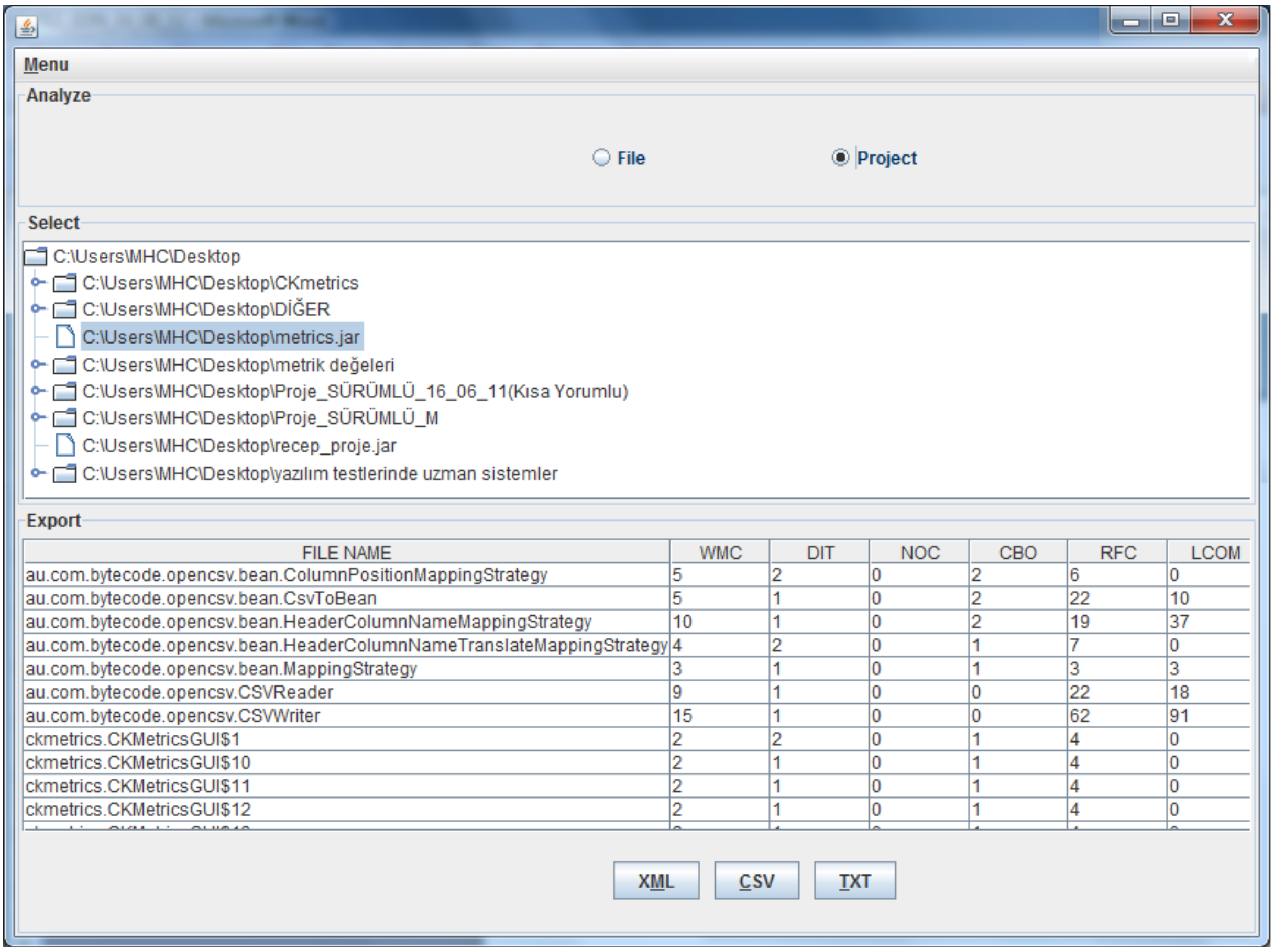

Şekil 2. Sisteme entegre edilen "Metrics" programından bir görünüm

## 3.2. Uzman Sistem Ünitesi

Uzman sistemler ünitesi, bilgi tabanı, çıkarım mekanizması ve raporlama ünitesi bileşenlerini içermektedir.

### 3.2.1. Bilgi tabanı

Bilgi tabanı, bir konuda bir veya birden çok uzmanın bilgilerinin bir araya getirilmesiyle oluşmaktadır (Başak vd., 2008). Bilgi tabanı, veri tabanı ve kural tabanını içermektedir.

*Veri Tabanı:* Problemin o andaki durumunu anlatan gerçekler ve belirli bir ana kadar elde edilmiş nitelik-değer çiftlerinden oluşmaktadır (Şahin ve Börklü, 2008). Yapılan çalışmada veriler, Delphi 7 programlama dili eklentisi olan Paradox veritabanı kullanılarak kaydedilmektedir. Kullanıcı isterse yapılan işlemleri kaydedebilmekte, isterse eski verileri geri çağırıp üzerinde istediği değişiklikleri yapabilmektedir.

*Kural Tabanı:* Değerlendirmek için kullanılan kuralların tutulduğu birimdir. Bu birimde, CK metrik kümesini değerlendirmek için, literatürde kabul gören ve kullanılan eşik değer aralıkları çalışmanın kural tabanını oluşturmaktadır (Erdemir vd., 2008; Chidamber ve Kemerer, 1994; Çatal ve Diri, 2008a; Chidamber ve Kemerer, 1991; Calp, 2011; Thirugnanam ve Swathi, 2010; Mccabe, 2012; Kaur vd., 2007).

Oluşturulan kural tabanı, "Bilginin 'eğer-o halde' kuralıyla sunulması" yöntemi ile kullanılmıştır. Uzman modülün kural temelini oluşturan metriklere ilişkin eşik değerler Çizelge 1'de gösterilmektedir.

Çizelge 1. Kuralları oluşturmada kullanılan eşik değerler (Calp, 2011b)

| | **Metrikler** | | | | | |
|---|---|---|---|---|---|---|
| | **SAMS** | **KAD** | **ASS** | **NSB** | **STMS** | **MUE** |
| **Eşik Değerleri** | 0-15 | 0-7 | 0-6 | 0-5 | 0-55 | 0-25 |

Çalışmanın kural tabanı, "Otomatik Değerlendirme" kısmında (eşik değerleri esas alarak) onsekiz adet kuraldan meydana gelirken, "El ile Değerlendirme" için eşik değerlerin kullanıcı tarafından değiştirilmesiyle yirmidört adet kuraldan oluşmaktadır.
Geliştirilen uzman sistemde, "Otomatik Değerlendirme" kısmı için eşik değerler kullanılarak oluşturulan kurallar Çizelge 2'de verilmiştir.

Çizelge 2. "Otomatik Değerlendirme" için oluşturulan kurallar

| | **Metrikler** | | | | | |
|---|---|---|---|---|---|---|
| | **SAMS** | **KAD** | **ASS** | **NSB** | **STMS** | **MUE** |
| | = 0 | = 0 | = 0 | = 0 | = 0 | = 0 |
| **Eğer** | 0-15 | 0-7 | 0-6 | 0-5 | 0-55 | 0-25 |
| | >15 | >7 | >6 | >5 | >55 | >25 |

Çalışmanın "El ile değerlendirme" kısmında kullanıcı tarafından oluşturulan kurallar ise Çizelge 3'te verilmiştir.

Çizelge 3. "El ile Değerlendirme" için oluşturulan kurallar

| | **Metrikler** | | | | | |
|---|---|---|---|---|---|---|
| | **SAMS** | **KAD** | **ASS** | **NSB** | **STMS** | **MUE** |
| | =0 | =0 | =0 | =0 | =0 | =0 |
| | 0-belirlenen ilk değer | 0-belirlenen ilk değer | 0-belirlenen ilk değer | 0-belirlenen ilk değer | 0-belirlenen ilk değer | 0-belirlenen ilk değer |
| **Eğer** | belirlenen ilk değer - belirlenen son değer | belirlenen ilk değer - belirlenen son değer | belirlenen ilk değer - belirlenen son değer | belirlenen ilk değer - belirlenen son değer | belirlenen ilk değer - belirlenen son değer | belirlenen ilk değer - belirlenen son değer |
| | >son değer | >son değer | >son değer | >son değer | >son değer | >son değer |

Sistemde, temel anlamda toplamda kırk iki adet kural bulunmaktadır. Bu kurallar kullanılarak geliştirilen uzman sistemin çıkarım mekanizması birimi ile, yazılımların iç kalite özellikleri hakkında birtakım çıkarımlar yapılmaktadır.

### 3.2.2. Çıkarım Mekanizması

Çıkarım mekanizması ise kurallara göre verileri işleyerek anlamlı çıkarımların yapılmasını sağlamaktadır. Çıkarım mekanizması, kuralları yorumlamada iki farklı arama metodu kullanmaktadır. Bunlar; ileri ve geriye zincirleme metotlarıdır. Bu çalışmada, bilinen verilerden başlayarak uygun kural bulununca ilgili kuralın mevcut şartlarını tahmin etmeyi temel alan “ileri zincirleme” metodu kullanılmıştır. Çizelge 4'te, uygulamada oluşturulan kural ve çıkarımlara, düşük, normal ve yüksek durumları için birer örnek verilmiştir.

Çizelge 4. Kural ve çıkarım örnekleri

<table>
<tr><th>Kural No</th><th rowspan="4">Eğer</th><th>Kural</th><th rowspan="4">O halde</th><th>Çıkarım</th></tr>
<tr><td>1</td><td>KAD değeri 5 ise;</td><td>Kalıtım Ağacının Derinliği: İstenen Aralıkta<br>Kod Hata Olma İhtimali: Düşük<br>Bakım, Onarım ve Test Faaliyetleri: Az<br>Kalite Düzeyi: Yüksek<br>Anlaşılırlık: Yüksek<br>Test Edilebilirlik: Yüksek<br>Yeniden Kullanılabilirlik: Yüksek<br>Karmaşıklık: Düşük</td></tr>
<tr><td>2</td><td>SAMS değeri 18 ise;</td><td>Sınıflardaki Metot Sayısı: Yüksek<br>Kod Hata Olma İhtimali: Yüksek<br>Bakım, Onarım ve Test Faaliyetleri: Çok<br>Kalite Düzeyi: Düşük<br>Anlaşılırlık: Düşük<br>Dayanıklılık: Düşük<br>Yeniden Kullanılabilirlik: Düşük<br>Karmaşıklık: Yüksek</td></tr>
<tr><td>3</td><td>NSB değeri 1 ise;</td><td>Bağımlılık Düzeyi: Çok Düşük<br>Modüler Tasarım: Çok Düşük<br>Kod Hata Olma İhtimali: Çok Düşük<br>Bakım, Onarım ve Test Faaliyetleri: Çok Az<br>Kalite Düzeyi: Çok Düşük<br>Anlaşılırlık: Yüksek<br>Yeniden Kullanılabilirlik: Çok Düşük<br>Karmaşıklık: Çok Düşük</td></tr>
</table>

### 3.2.3. Raporlama Ünitesi

Geliştirilen uzman değerlendirme modülünün yapısı içerisindeki uzman sistem ünitesinin elde ettiği sonuçların kullanıcılara iletilmek üzere raporlandığı bölümdür. Bu ünite, CK'nın daha önce elde edilen sayısal metrik bulgularının değerlendirmeleri ile ilgili kapsamlı bir rapor verecek şekilde tasarlanmıştır (Şekil 3).

**DEĞERLENDİRME SONUÇLARI**

ELDE EDİLEN BULGULAR WMC METRİĞİ AÇISINDAN İNCELENDİĞİNDE,
EN KÜÇÜK değere sahip sürüm/ler : SÜRÜM-5 (0)
EN BÜYÜK değere sahip sürüm/ler : SÜRÜM-2 ve SÜRÜM-3 (5)
Kod Kalitesi Bakımından;
SÜRÜM-2 ve SÜRÜM-3 : En Karışık ve En Düşük
SÜRÜM-5 : En Az Karışık ve En Yüksek
Geliştirilmesi, Bakım-Onarımı ve Test Faaliyetleri Bakımından Harcanak Zaman ve İş Gücü;
SÜRÜM-2 SÜRÜM-3 : En Çok
SÜRÜM-5 : En Az

Şekil 3. Raporlama ekranından bir görünüm

### 3.3. Uzman Modülün Uygulaması

Öncelikle, geliştirilen yazılıma "Metrics" programı entegre edilmiştir. Bu eklenti, nesne yönelimli java sınıf veya .jar dosyalarının sayısal metrik değerlerini kullanıcıya sunmaktadır. Bu çalışma, "Metrics" programı kullanılarak elde edilen sayısal bulguları "Sınıf veya Proje Değerlendirme" ve "Sürümler Arası Değerlendirme" olmak üzere iki durumda değerlendirecek şekilde geliştirilmiştir. Geliştirilen uzman modüle ilişkin aktivite diyagramı Şekil 4'te gösterilmektedir.

Uzman modülün "Sınıf veya Proje Değerlendirme" başlığı altında hem otomatik hem de el ile değerlendirme yapılabilmektedir. "Sınıf veya Proje" başlığı altında bulunan "Otomatik" alt menü içerisindeki metrik değerlendirmelerde kurallar, modülün gerçekleştirimi sırasında önceden veritabanına girilmiştir. Kullanıcı, istediği durumda bu kuralları görebilmekte ve ekleme veya değişiklik yapabilmektedir. Sonuç olarak geliştirilen sistem, kullanıcı tarafından çizelgeye aktarılan sayısal metrik bulguları otomatik olarak değerlendirmektedir (Çizelge 5).

Çizelge 5. Otomatik değerlendirme için oluşturulan çizelge örneği

| SINIFLAR | WMC | DIT | NOC | CBO | RFC | LCOM |
|---|---|---|---|---|---|---|
| moreUnit.actions.CreateTestMethodE | 4 | 1 | 0 | 6 | 9 | 4 |
| moreUnit.actions.CreateTestMethodH | 4 | 1 | 0 | 11 | 14 | 4 |
| moreUnit.actions.JumpAction | 4 | 1 | 0 | 5 | 7 | 4 |
| moreUnit.decorator.UnitDecorator | 4 | 0 | 0 | 17 | 21 | 6 |
| moreUnit.elements.ClassTypeFacade | 9 | 0 | 0 | 12 | 34 | 36 |
| moreUnit.elements.EditorPartFacade | 8 | 1 | 0 | 17 | 27 | 22 |
| moreUnit.elements.JavaProjectFacad | 5 | 1 | 0 | 13 | 29 | 0 |

Ancak, "El ile Değerlendirme" başlığı altındaki sayfada kural oluşturma modülü sayesinde yazılım bilgisine sahip olmadan da yeni kural belirleme işlemleri kolaylıkla yapılabilmektedir. Bu şekilde kullanıcılar metrik aralıkları için farklı eşik değer aralıkları oluşturabilmektedir.

Bu durum esneklik sağlamakta ve ileriki süreçte gelişen veya değişen metrik eşik değerleriyle ilgili bir güncelleme yapılabilmektedir. Çalışmada, metrik tipleri ve metrik aralıkları kullanıcının isteğine göre belirlenip bu tip ve aralıklar dikkate alınarak değerlendirme yapılmaktadır (Şekil 5).

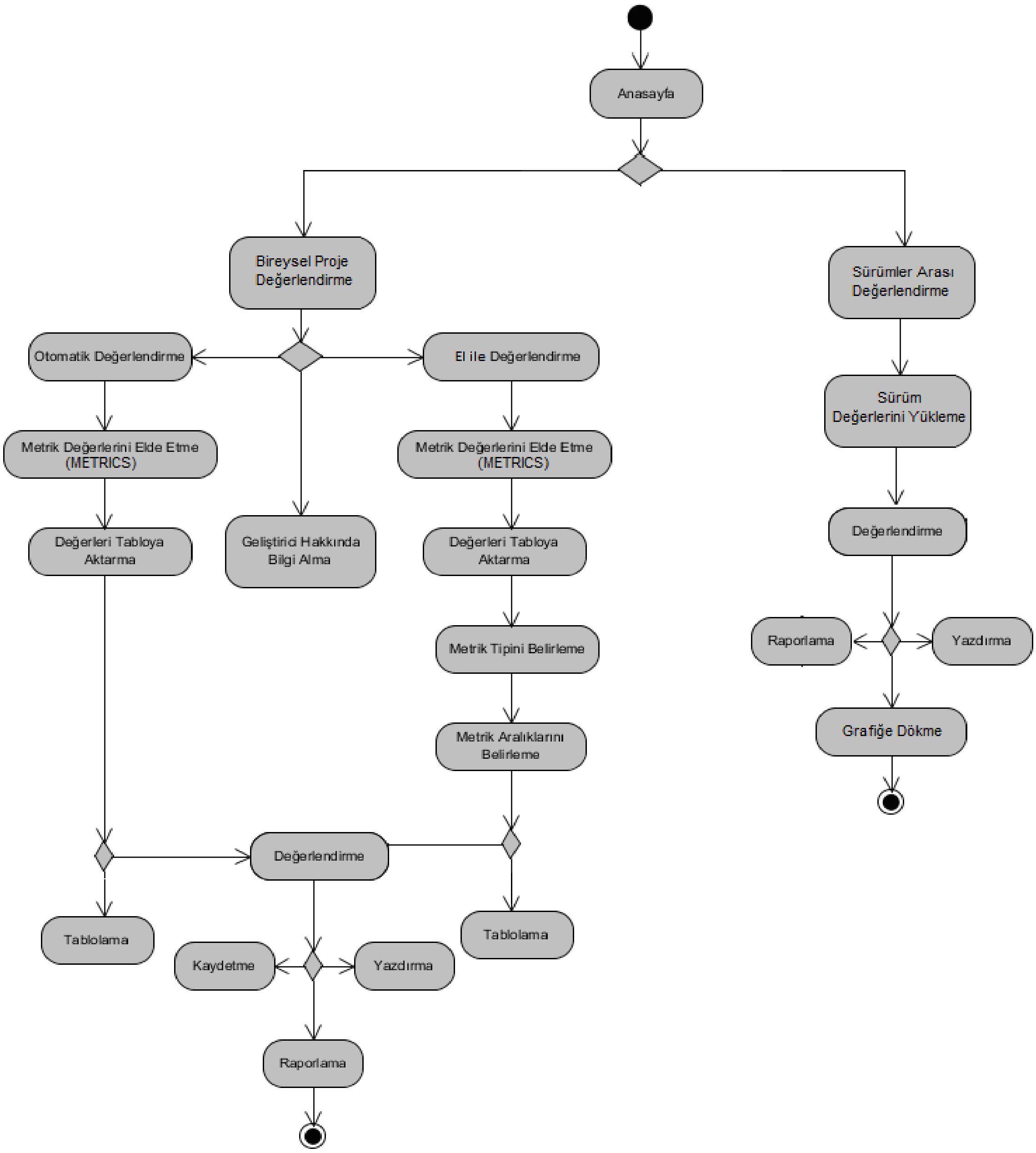

Şekil 4. Aktivite diyagramı

"Sürümler Arası Değerlendirme" yönteminde ise, sürümlerin metrik değerleri karşılaştırılarak değerlendirme yapılmaktadır. Bu yöntemde, aynı metrik grubunun ölçülecek olan yazılımın sürümleri üzerinde uygulanarak değerlendirme yapılmaktadır. Başka bir ifadeyle, önceki yazılım sürümlerin metrik değerleri ve kusur bilgileri kullanılarak bir model ortaya çıkarılmaktadır. Ardından yeni sürümdeki metrik değerleri ve kusur bilgileri ile karşılaştırılarak kalite düzeyi yorumlanmaktadır. Şekil 6'da bu durumu gösteren örnek bir uygulama ekranı verilmiştir.

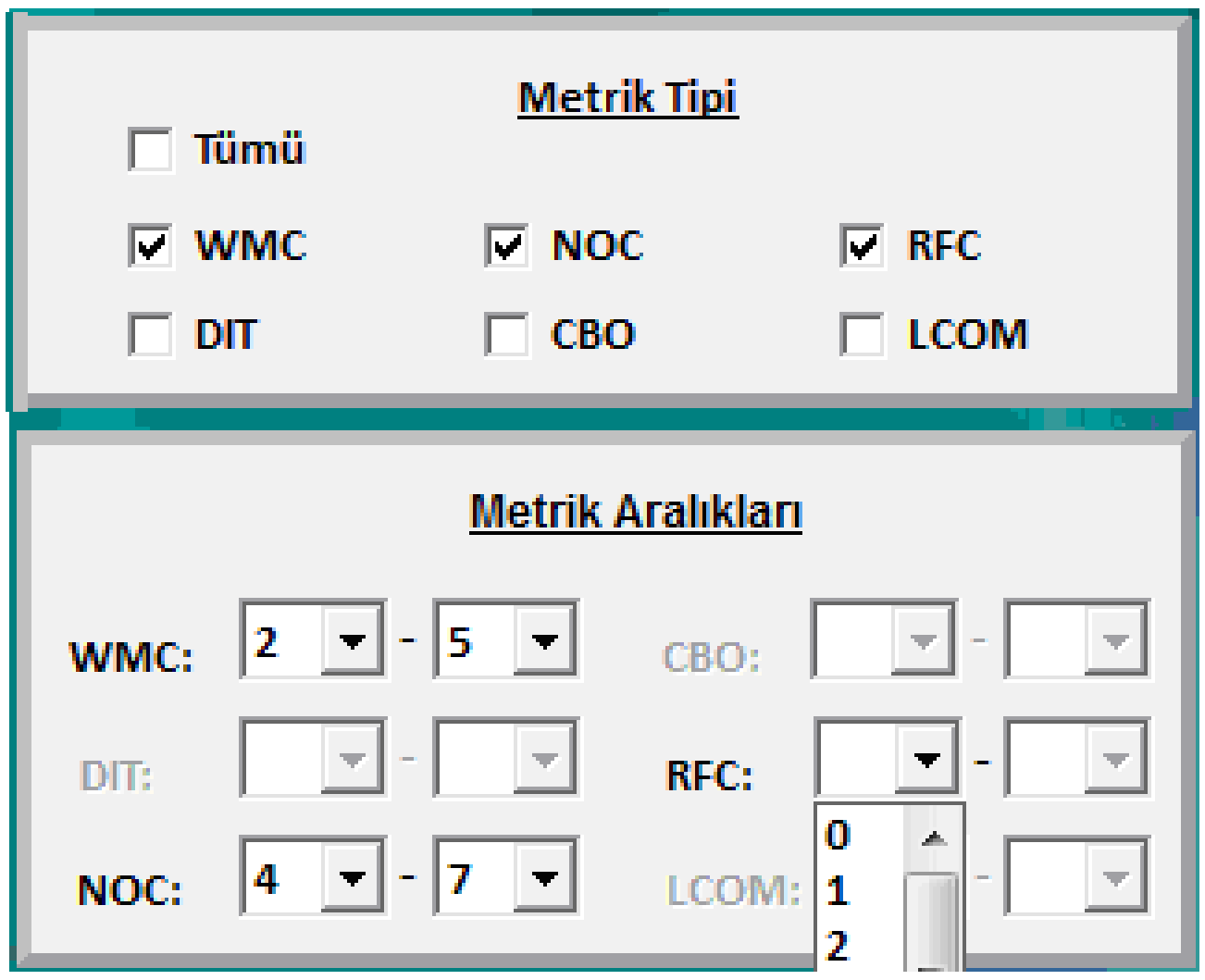

Şekil 5. Metrik tipi ve aralığının seçilmesi

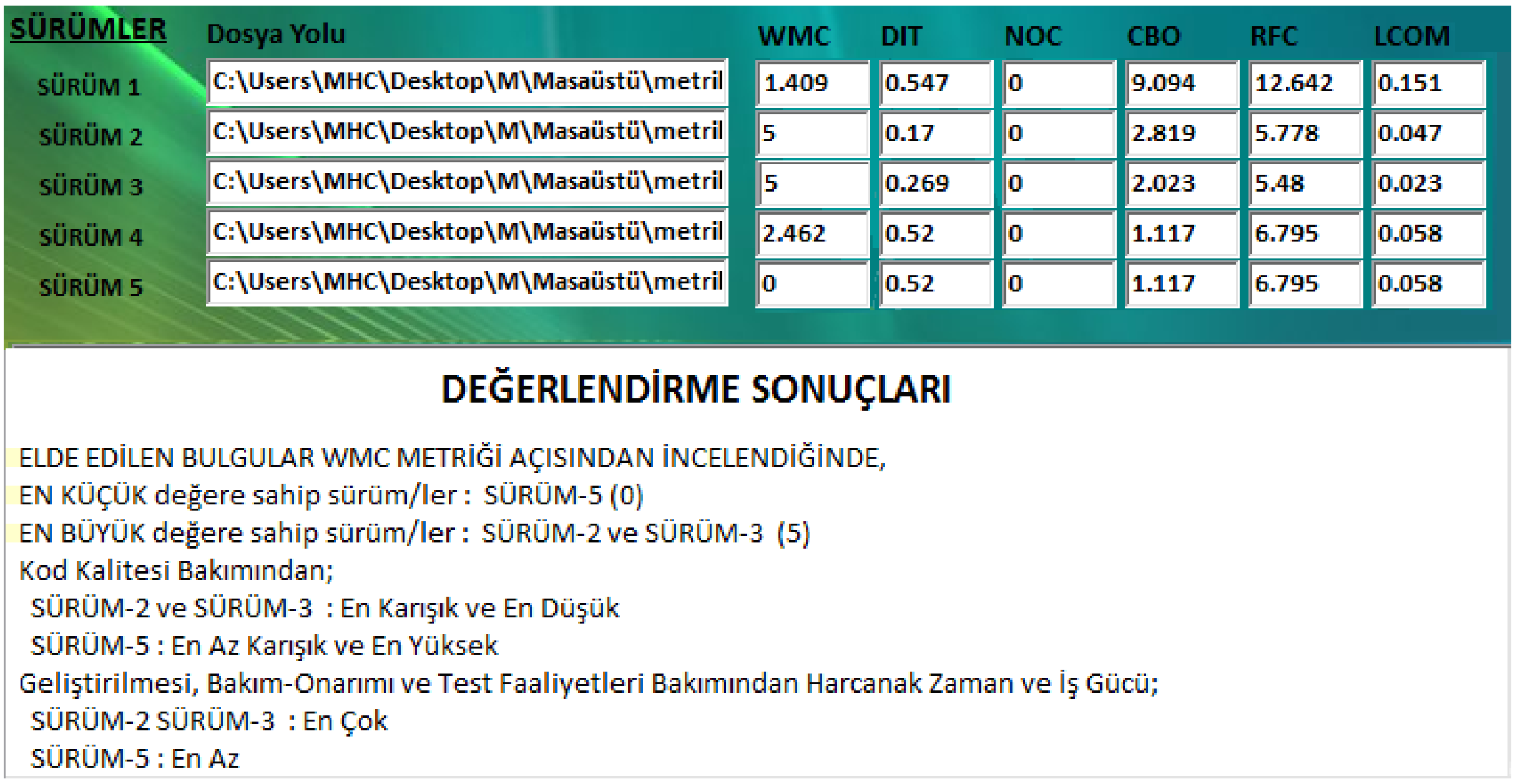

Şekil 6. Sürümler arası değerlendirme sayfasından bir görünüm

Tüm bu sonuçların metin dosyası halinde raporlanması da gerçekleştirilebilmektedir. Aynı zamanda elde edilen bu rapor, geliştirilen yazılımın "Dosya" menüsü kullanılarak düzenlenebilmekte, Microsoft Word (.doc) ve metin dosyası (.txt) formatlarında kaydedilebilmekte veya yazdırılabilmektedir.

Son olarak, "Metrics" programı ile elde edilen verilerin sürümlere göre metrik değerlerini bir arada görmek ve net bir şekilde ortaya koymak amacıyla isteğe bağlı olarak elde edilen metrik bulguları grafiğe dökülebilmektedir. Şekil 7'de, Şekil 6'da verilen değerler kullanılarak oluşturulan örnek bir grafik görünümü mevcuttur.

## 4. SONUÇ VE ÖNERİLER

Günümüzde yazılım hatalarının sebeplerini ortadan kaldıracak veya bu hataların oluşmasını engelleyecek çalışmalar yapılmaktadır. Ancak tüm bu çalışmalar yapılmasına rağmen yazılımlardaki veya projelerdeki hatalar tam manasıyla engellenememektedir.

Bu çalışmada, özellikle nesne yönelimli yazılım metrikleri ve değerlendirilmesi araştırılmıştır. Ayrıca, nesne yönelimli yazılımlar temelde sınıflandırma mantığıyla geliştirildiği için, sözkonusu yazılımların sınıf metriklerini değerlendiren bir uzman modül gerçekleştirilmiştir. Geliştirilen uzman modül aracılığı ile .jar uzantılı bir dosyanın veya projenin sürümlerinin metrik bulguları değerlendirilmektedir. Sözkonusu metrikler literatürde yaygın olarak kullanılan C&K'in metrik kümesidir. Değerlendirmede uzman sistem yaklaşımı tercih edilmiştir. Bu modülün "Sınıf veya Proje Değerlendirilmesi" kısmında uzman sistemlerin bilgi sunma kurallarından olan "Eğer-O halde" kural yapısı kullanılmıştır. Değerlendirme yöntemi olarak, "*özelliğe bağlı metrik ölçümü ve yorumlanması*" yöntemine başvurulmuştur. "Sürümler Arası Değerlendirme" kısmında ise, yine metrik değerlendirme yöntemlerinden olan "*daha önceki sürüm veya sürümler ile yeni geliştirilen sürümün metrik bulguları karşılaştırılması*" yöntemine başvurularak geliştirilen yazılımın iç kalitesi hakkında yorumlar yapılabilmektedir. Buna ilaveten geliştirilen modül, metrik kümesinden istenildiği kadar metrik işleme dahil edilerek kullanıcıya metrikleri bir arada inceleyebilme imkânı sunmaktadır. Sonuç itibariyle bu çalışmada, uzman sistemler aracılığıyla yazılım metrik değerlendirilmesi ve bunun sonucu olarak da yazılım kalite değerlendirilmesi konularında bir otomasyon sağlanmıştır.

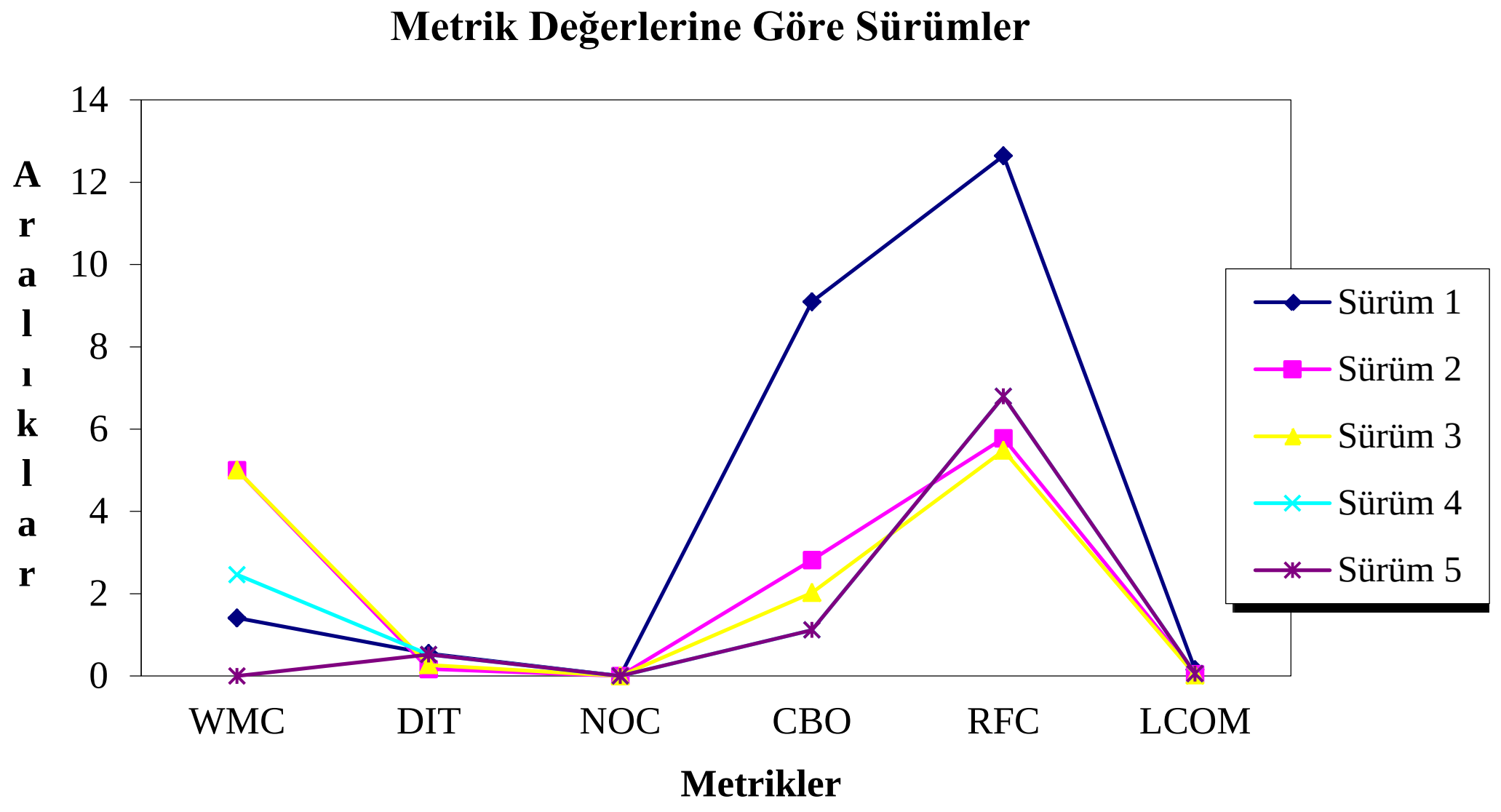

Şekil 7. Sürümlerin metrik değerlerine göre grafiksel gösterimi

Ancak, ileriki çalışmalarda metriklerin değerlendirilmesinde, belirlenen metriklerin ürettiği değerler ile aynı amaca hizmet eden diğer metriklerin ürettiği değerler arasındaki ilişki ölçülerek yapılan değerlendirme yöntemi de kullanılabilir. Başka bir ifadeyle, farklı metrik kümeleri arasındaki ilişki kullanılarak geliştirilen yazılımın kalitesi ortaya koyulabilir.

Projenin bundan sonraki aşamasında aynı anda birden fazla metriği değerlendiren, metriklerin birbiriyle ilişkisini ortaya koyan ve bu ilişki sonucu sistemin kalitesi hakkında değerlendirme yapabilen bir sistem geliştirilmesi planlanmaktadır. Bu değerlendirmede, kolay görülemeyecek hataların belirlenmesi ve daha sağlıklı sonuçlar elde edilmesi açısından diğer yapay zeka tekniklerinin kullanılması da düşünülmektedir.

Sonuç olarak, yapılan çalışmada nesne yönelimli tasarım metrik değerlendirilmesi ve bunun sonucu olarak da yazılım kalite değerlendirilmesi konularında bir otomasyon geliştirilerek beyaz kutu test sürecine katkı sağlayacak alt yapı hazırlanmıştır.

**KAYNAKLAR**

A. L. Baroni, S. Braz, F. B. Abreu (2002): "Using OCL to Formalize Object-Oriented Design Metrics Definitions", 6th International ECOOP Workshop on Quantitative Approaches in Object-Oriented Software Engineering, Malaga, İspanya.

V. R. Basili, L. C. Briand, W. Melo (1996): "A Validation of Object-Oriented Design Metrics as Quality Indicators," IEEE Trans. on Software Engiıneering, Cilt 22, No. 10.

H. Başak, İ. Şahin, M. Gülen (2008): "İnsansız Hava Aracı Kazalarının Önlenmesi İçin Uzman Sisteme Dayalı Risk Yönetim Modeli", Teknoloji, Cilt 11, No. 3, sf. 187-200.

D. Bellin, M. Tyagi, M. Tyler (1999): "Object-Oriented Metrics: An Overview", Bilgisayar Mühendisliği Bölümü, Department,North Carolina, AT State Universitesi, Greensboro, Nc 27411-0002.

L. C. Briand, S. Morasca, V. R. Basili (1996): "Property-Based Software Engineering Measurement", IEEE Trans. on Software Engineering, Cilt 22, No. 1.

F. B. Abreu, R. Esteves, M. A. Goulao (1996): "The Design of Eiffel Programs: Quantitative Evaluation Using the MOOD Metrics", actas de TOOLS'96 (Technology of Object Oriented Languages and Systems), Santa Barbara, CA, EUA, Julho.

M. H. Calp, N. Arıcı (2011): "Nesne Yönelimli Tasarım Metrikleri ve Kalite Özellikleriyle İlişkisi", Gazi Üniversitesi Teknik Eğitim Fakültesi Politeknik Dergisi, Cilt: 14, Sayı: 1, sf. 9-14.

M. H. Calp (2011): "Nesne Yönelimli Yazılım Testi ve Metrik Kümesi Değerlendiren Uzman Modülün Gerçekleştirilmesi", Yüksek Lisans Tezi, Gazi Üniversitesi, Bilişim Enstitüsü, Ankara.

N. Chapin, J. Hale, K. Khan, J. Ramil (2001): "Type of software evolution and software maintenance". Journal of Software Maintenance and Evolution.

S. R. Chidamber, C. F. Kemerer (1991): "Towards A Metric Suite for Object-Oriented Design", Proceedings : OOPSLA'91, Phoenix, sf.197-211.

S. R. Chidamber, C. F. Kemerer (1994): "A Metrics Suite for Object-Oriented Design", IEEE Transactions On Software Engineering, Cilt 20, No. 6, sf. 482-491.

Coverity (2012): "Coverity Analysis", Çevrimiçi: http://www.coverity.com/, Son Erişim Tarihi: 13.04.2012.

Ç. Çatal, B. Diri (2008b): "A Fault Prediction Model with Limited Fault Data to Improve Test Process", Lecture Notes in Computer Science 5089, Springer-Verlag, sf. 244-257.

Ç. Çatal, B. Diri (2007): "Software Fault Prediction with Object Oriented Metrics Based Artifical Immune Recognition System", Lecture Notes in Computer Science 4589, Springer-Verlag, sf. 300-314.

Ç. Çatal, B. Diri (2008a): "Yazılım Metriklerini Kullanarak Düşük Kaliteli/Yüksek Kaliteli Modüllerin Otomatik Tespiti", Yazılım Kalitesi ve Yazılım Geliştirme Araçları Sempozyumu, İstanbul, 1, sf. 8-9.

U. Erdemir, U. Tekin, F. Buzluca (2008): "Nesneye Dayalı Yazılım Metrikleri ve Yazılım Kalitesi", Yazılım Kalitesi ve Yazılım Geliştirme Araçları Sempozyumu, İstanbul, sf. 4.

Findbugs (2012):, "Find Bugs in Java Programs", Çevrimiçi: http://findbugs.sourceforge.net/index.html, Son Erişim Tarihi:13.04.2012.

R. Harrison, S. Counsell (1998): "Theoretical Validation and Empirical Evaluation of OO Design Metrics", Declarative Systems and Software Engineering Group.

J. P.Kaur, A. Verma, S. Thapar (2007): "Software Quality Metrics for Object-Oriented Environments", Proceedings of National Conference on Challenges & Opportunities in Information Technology, sf.13-16.

V. Laing, C. Coleman (2001): “Principal Components of Orthogonal OO Metrics", Software Assurance Tecknoloji Merkezi.

M. Lanza, S. Ducasse (2001): “A Categorization of Classes based on the Visualization of their Internal Structure: The Class Blueprint”, Proceedings of OOPSLA 2001, sf. 300 - 311, ACM Press.

M. Lanza, S. Ducasse (2003): “Polymetric Views-A Lightweight Visual Approach to Reverse Engineering”; In IEEE Trans. on Software Engineering, Cilt. 29, No. 9, sf. 782 - 795, IEEE CS Press.

M. Lanza, R. Marinescu (2006): “Object-Oriented Metrics in Practice”; Springer; ISBN: 3-540-24429-8.

R. Marinescu (2004): “Detection strategies: Metrics-based rules for detecting design flaws”, ICSM'04, Los Alamitos, IEEE Computer Society Press, sf. 350-359.

Mccabe Software (2012): “Using Code Quality Metrics in Management of Outsourced Development and Maintenance”, Çevrimiçi: http://www.mccabe.com/pdf/mccabecodequalitymetrics-outsourceddev.pdf, Son Erişim Tarihi:13.04.2012.

Metrics 1.3.6 (2012): Çevrimiçi: http://metrics.sourceforge.net/, Son Erişim Tarihi:13.04.2012.

J. F. Muratore, M. P. Demasie (1991): “Artificial Intelligence and Expert Systems In-Flight Software Testing”, Digital Avionics Systems Conference, 1991 Proceedings, IEEE/AIAA 10th, Los Angeles, Amerika, sf. 416-419.

V. V. Nabiyev (2005): “Yapay Zeka”, Seçkin Yayıncılık, Ankara.

PMD (2012):, Çevrimiçi: http://pmd.sourceforge.net/, Son Erişim Tarihi:13.04.2012.

R. S. Pressman (2005): “Software Engineering: A Practitioner’s Approach”, Mc Graw Hill, Singapur.

R. Pusala (2005): “Operational Excellence Through Efficient Software Testing Metrics”, Eurostar 2005: European Conference On Software Testing Infosys, Almanya, sf. 2-3.

C. L. Ramsey, V. R. Basili (1986): “An Evaluation of Expert Systems for Software Engineering Management”, Bilgisayar Mühendisliği Bölümü, Maryland College Park Universitesi, Teknik Rapor: TR 1708, IEEE Trans. on Software Engineering, sf. 747–759.

C. M. Robert (1995): “Object Oriented Design Quality Metrics: An Analysis of dependencies”, ROAD, Cilt 2, No. 3.

M. Sarker (2005): “An Overview of Object Oriented Design Metrics”, Master Thesis: Bilgisayar Mühendisliği Bölümü, Umea Universitesi, İsveç.

B. H. Sellers (1996): “Object-Oriented Metrics-Measures of Complexity, série:The Object-Oriented Series”, Prentice Hall PTR, ISBN 0-13-239872.

F. T. Sheldon, K. Jerath, H. Chung (2002): “Metrices for Maintainability of Class Inheritance Hierarchies”, Journal of Software Maintenance and Evaluation.

İ. Şahin, H. R. Börklü (2008): “2B Görünüşlerden Otomatik Katı Modeller Oluşturmada Uzman Bir Yaklaşım”, Mühendislik Bilimleri Dergisi, Cilt 14, No. 2, sf. 111-123.

Y. Tao (1995): “Using Expert Systems to Understand Object-Oriented Behaviour”, 26th SISCSE Technical Symposium on Computer Science Education.

M. Thirugnanam, J. N. Swathi (2010): “Quality Metrics Tool for Object Oriented Programming”, International Journal of Computer Theory and Engineering, Cilt 2, No. 5, sf. 1793-8201.

Z. Xu, T. M. Khoshgoftaar, E. B. Allen (2003): “Application of Fuzzy Expert Systems in Assessing Operational Risk of Software”, Information and Software Technology, Cilt 45, No. 7, sf. 373-388.

Z. Xu, T. M. Khoshgoftaar, K. Gao (2005): “Application of Fuzzy Expert System in Test Case Selection for System Regression Test”, Proceedings of IEEE International Conference on Information Reuse and Integration, IEEE Computer Society, Las Vegas.